\documentclass[reprint,superscriptaddress,amsmath,amssymb,aps,prab,]{revtex4-2}

\usepackage{graphicx}
\usepackage{subcaption}
\usepackage{dcolumn}
\usepackage{bm}
\usepackage{hyperref}
\hypersetup{hidelinks}
\usepackage{tikz}
\usepackage{lipsum}
\usetikzlibrary{arrows.meta, positioning, shadows, patterns}
\definecolor{sagebox}{RGB}{155,173,155}
\definecolor{scfield}{RGB}{230,145,20}
\definecolor{appfield}{RGB}{25,90,180}
\definecolor{cathfill}{RGB}{225,218,200}
\definecolor{cathedge}{RGB}{70,90,120}

\begin{document}

\preprint{APS/123-QED}

\title{Building a Start-to-End Model of the CESR Injector Linac}

\author{Ryland Goldman}
\affiliation{Cornell Laboratory for Accelerator-Based Sciences and Education, Ithaca, New York 14853, USA}
\affiliation{University of California, Los Angeles, California 90095, USA}
\author{Adam Bartnik}
\affiliation{Cornell Laboratory for Accelerator-Based Sciences and Education, Ithaca, New York 14853, USA}
\author{Jared Maxson}
\affiliation{Cornell Laboratory for Accelerator-Based Sciences and Education, Ithaca, New York 14853, USA}

\date{\today}

\begin{abstract}
The CESR injector linac supplies positrons to the Cornell Electron Storage Ring in six stages: a thermionic cathode, a 150~kV DC gun, a prebuncher, four electron linac sections, a tungsten positron converter, and four positron linacs. We present a start-to-end model of the full chain, built in Python from open-source codes. WarpX is used for the space-charge-dominated cathode, gun, prebuncher, and electron linacs, Geant4 for positron production, and Impact-T for the positron linacs. The beam reaches the converter at 146~MeV and the recaptured positrons are accelerated to ${\sim}250$~MeV. Stage configurations are written in YAML to allow algorithmic adjustment by Xopt for ML optimization.
\end{abstract}

\maketitle

\section{Introduction}

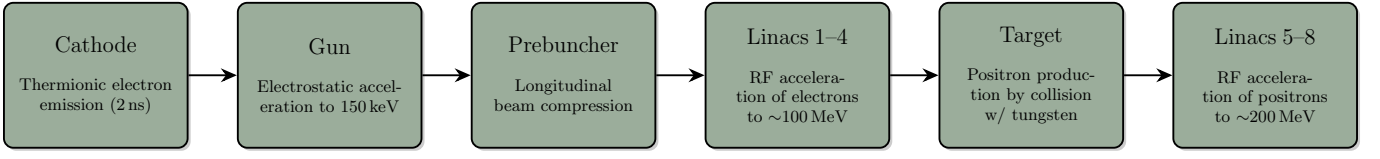
\begin{figure*}[t]
\centering
\resizebox{\textwidth}{!}{%
\begin{tikzpicture}[
    node distance=9mm,
    box/.style={
        rectangle, rounded corners=5pt,
        draw=black, line width=1pt,
        fill=sagebox,
        drop shadow={shadow xshift=1.6pt, shadow yshift=-1.6pt, opacity=0.4},
        text width=30mm, align=center,
        minimum height=27mm, inner sep=6pt,
    },
    arr/.style={-{Stealth[length=3mm, width=2.6mm]}, line width=1pt},
]
 
\node[box] (cathode) {%
    {\large Cathode}\\[10pt]
    {\small Thermionic electron emission (2\,ns)}%
};
\node[box, right=of cathode] (gun) {%
    {\large Gun}\\[10pt]
    {\small Electrostatic acceleration to 150\,keV}%
};
\node[box, right=of gun] (prebuncher) {%
    {\large Prebuncher}\\[10pt]
    {\small Longitudinal beam compression}%
};
\node[box, right=of prebuncher] (linac14) {%
    {\large Linacs 1--4}\\[10pt]
    {\small RF acceleration of electrons to ${\sim}100$\,MeV}%
};
\node[box, right=of linac14] (target) {%
    {\large Target}\\[10pt]
    {\small Positron production by collision w/ tungsten}%
};
\node[box, right=of target] (linac58) {%
    {\large Linacs 5--8}\\[10pt]
    {\small RF acceleration of positrons to ${\sim}200$\,MeV}%
};
 
\draw[arr] (cathode)    -- (gun);
\draw[arr] (gun)        -- (prebuncher);
\draw[arr] (prebuncher) -- (linac14);
\draw[arr] (linac14)    -- (target);
\draw[arr] (target)     -- (linac58);
 
\end{tikzpicture}%
}
\caption{The six stages of the CESR injector linac.}
\label{fig:linac}
\end{figure*}

The Cornell Electron Storage Ring (CESR) is a 768~meter positron ring at Cornell University. Originally used for high-energy electron-positron collisions, it now powers the Cornell High Energy Synchrotron Source (CHESS)~\cite{chessu}. Positrons are produced for the synchrotron using a six-stage linac outlined in Fig.~\ref{fig:linac}.

Electrons are emitted in a 2~ns pulse from a thermionic cathode and accelerated electrostatically up to 150~keV. They then undergo bunching in two low-frequency RF cavities, are accelerated up to 150~MeV, and impact a tungsten target. This target produces an electromagnetic shower which includes positrons that are collimated and reaccelerated to around 250~MeV.

An operational chain currently exists as LinacSim using General Particle Tracer and written in 2013~\cite{linacsim,gpt}. Several attempts have been made to modernize the code through BMAD, however the input files are difficult to parse~\cite{bmad}. This work presents a version of the simulation using Python with three actively maintained open-source programs, WarpX, Geant4, and Impact-T~\cite{warpx, geant4, impactt}.

A possible use of the updated code is for AI optimization of the operating parameters. The program allows for Xopt to run a constrained non-dominated genetic sorting algorithm (CNSGA)~\cite{xopt}. It is single-threaded to minimize memory latency and data transfer overhead, however multithreading could be implemented relatively easily.

\section{Methods}
\begin{figure}[htbp]
  \centering
  \includegraphics[width=\linewidth]{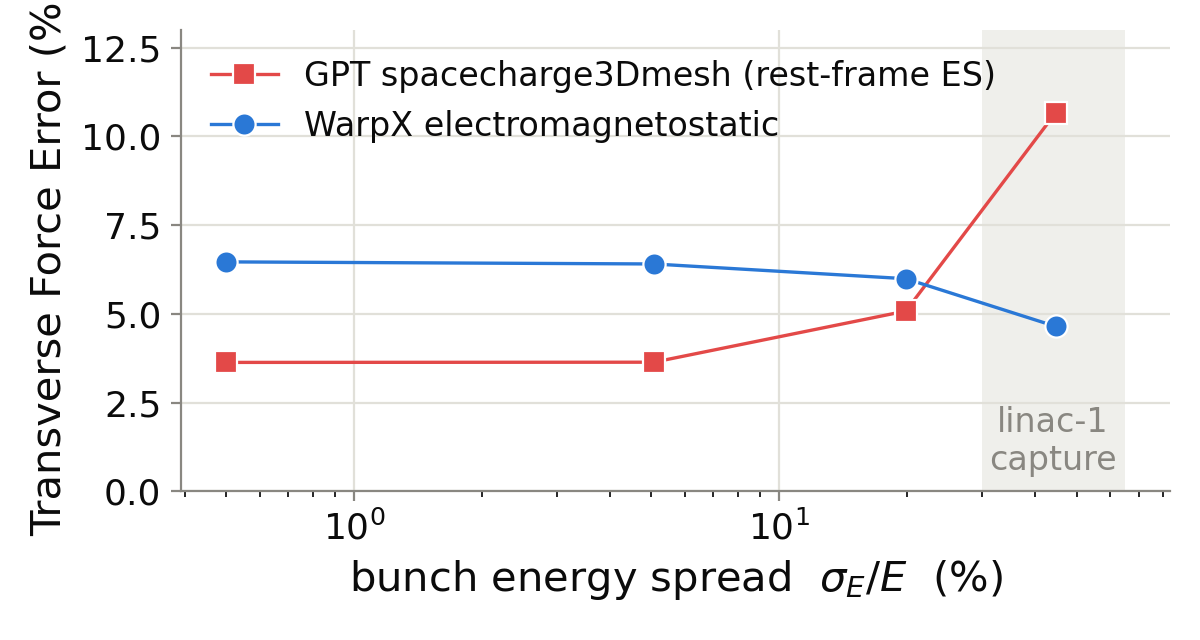}
  \caption{Transverse space-charge force error of GPT's rest-frame electrostatic mesh solver ({spacecharge3Dmesh}) and WarpX's electromagnetostatic solver versus bunch energy spread, measured against GPT's full $N$-body solver ({spacecharge3D}).}
  \label{fig:sc-solver-error}
\end{figure}

The electron portion is built using the WarpX particle-in-cell code using 2D azimuthally symmetric geometry. Space charge is enabled in the cathode, gun, prebuncher, and first linac using the electromagnetostatic solver, as GPT and other relativistic solvers approximate the calculation with electrostatics in the bunch rest frame (a poor approximation when the energy spread is high, as shown in Fig.~\ref{fig:sc-solver-error}). After the first linac, the mean kinetic energy is assumed to be high enough ($\langle\text{KE}\rangle=31.5$~MeV) such that the space charge effects can be ignored, since they are suppressed by a factor of $1/\gamma^2$.

The positron conversion process is simulated with Geant4 using the \verb|QGSP_BERT_EMZ| physics list~\cite{geant4}. To allow for better integration with the other beam components, the G4beamline wrapper is used~\cite{g4beamline}. This output is fed into Impact-T for the final positron acceleration.

No new fieldmaps were created in this work. Existing models created in Poisson Superfish~\cite{superfish} were used for the prebuncher and first linac section (based on the SLAC linac), however neither fieldmaps nor CAD models were available for the other linac sections. These other sections reused the first SLAC fieldmap, which is nonphysical but provides a good approximation. The positron capture optics also used previously generated BMAD fieldmaps found on CLASSE fileservers, which likely derive from earlier work on this injector~\cite{fromowitz}.

For ease of use, particle data is exchanged via the openPMD standard~\cite{openpmd}, fieldmaps are converted to HDF5 files, and configuration data is stored in YAML to be easily edited both by humans and computers. The particle count is upsampled to 50,000 after each stage due to beam losses throughout the linac.

\section{Cathode and Gun}
Electrons at CESR are created via thermionic emission from a 1425~K tungsten cathode and are accelerated via an electrostatic gun known as the Cornell High Intensity Linac Injector (CHILI) which operates at 150~kV DC~\cite{chili}. The emission is controlled by a grid bias which is pulsed for 2~ns to release the electrons~\cite{billing2000}. Computation for this stage is split into the emission calculation, which determines the initial particle distribution, and the electrostatic acceleration through the gun.

\subsection{Cathode Theory}

\begin{figure}[htbp]
    \centering
    \resizebox{\linewidth}{!}{
    \begin{tikzpicture}[
  >={Stealth[length=2.6mm,width=2mm]},
  efield/.style={->, scfield, line width=1.1pt},
  electron/.style={circle, fill=black!85, inner sep=1.1pt},
  font=\sffamily
]

\fill[cathfill] (0,0) rectangle (1.2,3.4);
\pattern[pattern=crosshatch, pattern color=black!12] (0,0) rectangle (1.2,3.4);
\draw[cathedge, line width=1.2pt] (0,0) rectangle (1.2,3.4);
\node[above, font=\sffamily\large] at (0.6,3.55) {Cathode, $V = 0$};

\draw[line width=2.2pt, dash pattern=on 7pt off 5pt, black!85] (7.2,-0.05) -- (7.2,3.45);
\node[above, font=\sffamily\large] at (7.2,3.55) {Mesh, $V = V_0$};

\draw[|<->|, black!70, line width=0.7pt] (-0.45,0) -- (-0.45,3.4)
  node[midway, left=1pt, font=\sffamily\small, rotate=90, anchor=south] {$16$~mm};
\draw[|<->|, black!70, line width=0.7pt] (1.25,3.15) -- (7.15,3.15)
  node[midway, above=1pt, font=\sffamily\small] {$d = 200\ \mu$m};

\foreach \y in {2.55, 1.85, 1.15, 0.45}
  \draw[efield] (1.25,\y) -- (2.15,\y);
\node[scfield, anchor=west, font=\sffamily\large] at (-0.45,-0.75) {Field from charges};

\foreach \p in {(2.45,2.75), (3.0,2.5), (2.55,2.2), (3.35,2.3), (3.95,2.45),
                (4.6,2.3), (2.5,1.75), (3.2,1.6), (3.85,1.7), (4.75,1.9),
                (2.55,1.25), (2.95,1.05), (3.9,1.4), (4.15,1.15), (4.3,0.8),
                (4.9,0.55), (4.6,0.35), (5.0,1.45), (5.3,2.1), (5.7,1.7)}
  \node[electron, label={[label distance=-1pt]right:{\footnotesize$e^-$}}] at \p {};

\draw[<-, appfield, line width=3.2pt] (4.1,-0.75) -- (7.15,-0.75);
\node[appfield, above=2pt, font=\sffamily\large] at (5.6,-0.68) {Applied field $E_a$};

\end{tikzpicture}
}
\caption{Schematic of the gridded thermionic cathode. Electrons emitted from the tungsten cathode are pulled across the 200~$\mu$m gap by a voltage $V_0$ from the mesh. The space-charge field of the electrons opposes the field, limiting the current density to the Child--Langmuir value $J_{CL}$.}
\label{fig:cathode-schematic}
\end{figure}
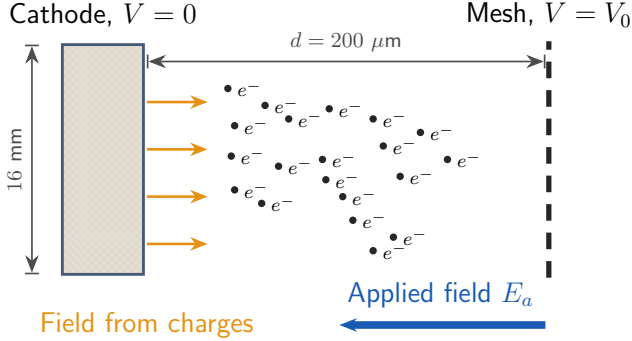

In a simplified version of the cathode, the system can be reduced to one dimension, as the diameter of the cathode is much greater than the cathode-grid separation $2R \gg d$. We consider the charges as infinitely wide sheets of charge density, each producing a uniform electric field, between two electrodes at a potential difference, as seen in Fig.~\ref{fig:cathode-schematic}. The total field in the region can be written as the superposition of a linearly increasing field from the potential difference, plus the field that the charged sheets would make if both conductors were grounded. For the latter, the field due to the $i$'th charge in this region is~\cite{linacsim}:
\begin{equation}
E_i = -\frac{z_i}{d}\frac{\sigma_i}{\epsilon_0} +
\begin{cases}
\sigma_i/\epsilon_0 & \text{left of the sheet,}\\
0 & \text{right of the sheet,}
\end{cases}
\label{eq:sheet-field}
\end{equation}
where $z_i$ and $\sigma_i$ are the position and charge density of sheet $i$, and $d$ is the electrode separation. With the sheets sorted in order of increasing $z$, the total field acting on the $n$'th sheet due to all the others is
\begin{equation}
E_{\text{total},n} = \sum_{i=1}^{n-1} E_i^R + \sum_{i=n+1}^{N} E_i^L,
\label{eq:sheet-total}
\end{equation}
where $E_i^L$ and $E_i^R$ denote Eq.~\ref{eq:sheet-field} evaluated to the left and right of sheet $i$. Splitting each term into its background and sheet parts, this can be rewritten as
\begin{equation}
E_{\text{total},n} = \sum_{i=1}^{N} E_i^{BG} + \sum_{i=1}^{N} E_i^{S} - E_n^{BG} - \sum_{i=1}^{n} E_i^{S},
\label{eq:sheet-cumsum}
\end{equation}
\begin{equation}
E_i^{S} = \frac{\sigma_i}{\epsilon_0}, \qquad E_i^{BG} = -\frac{z_i}{d}\frac{\sigma_i}{\epsilon_0},
\end{equation}
which yields the net longitudinal force on a sheet in that region.

As the charge from the previously-emitted sheets increases, it counters the applied voltage and the electric field immediately at the surface of the cathode drops. In steady-state, the self-limiting space charge emission follows the Child--Langmuir law~\cite{child1911,langmuir1913}:
\begin{equation}
J_{CL} = \frac{4\epsilon_0}{9}\sqrt{\frac{2e}{m_e}}\,\frac{V_0^{3/2}}{d^2}.
\label{eq:child-langmuir}
\end{equation}

\subsection{Gun Fieldmap}
\begin{figure}[htbp]
  \centering
  \includegraphics[width=\linewidth]{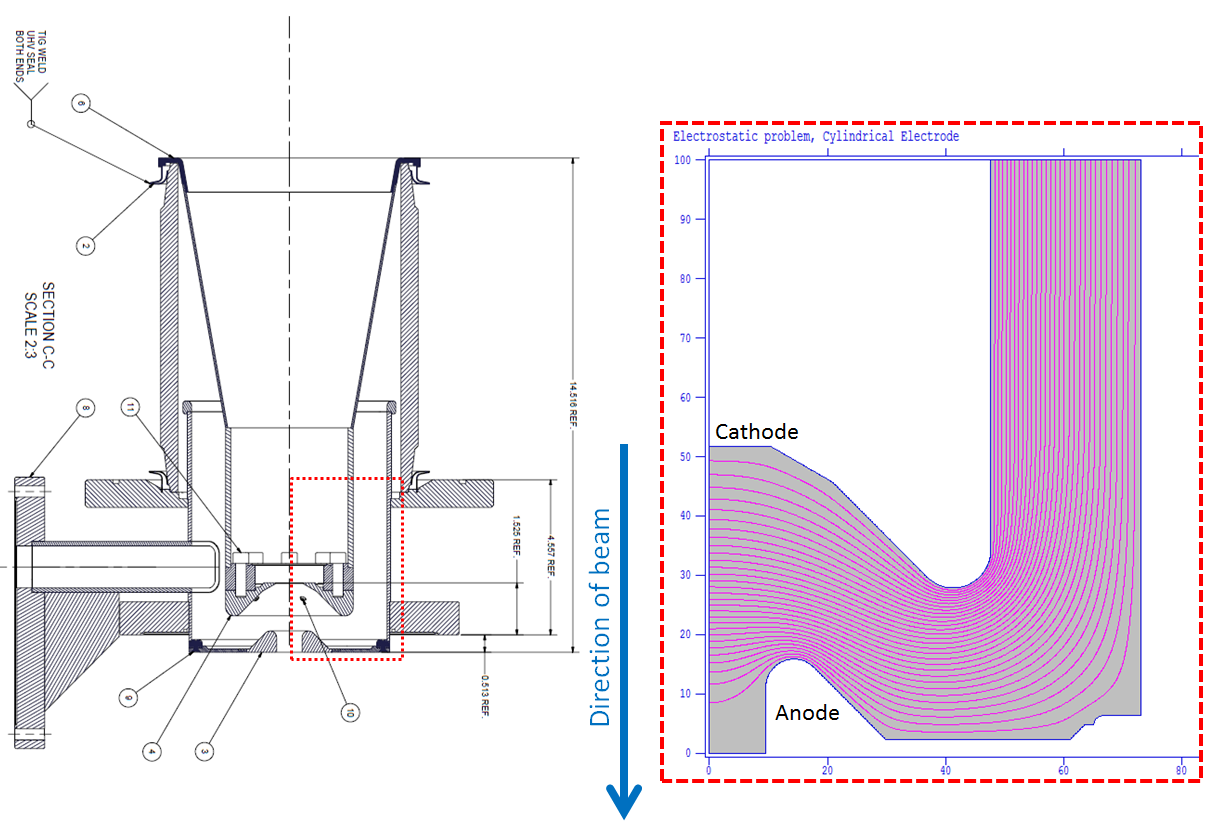}
  \caption{Left, drawing 6051-022 of CHILI gun~\cite{chiligun_dwg}; right, generated 2D fieldmap exported to WarpX (equipotentials shown).}
  \label{fig:gun-fieldmap}
\end{figure}

The fieldmap for the CHILI gun was created prior to this work using scans from the engineering drawings in Fig.~\ref{fig:gun-fieldmap}~\cite{linacsim,chiligun_dwg}. A cylindrically symmetric cathode and anode structure was used to build the model in Superfish, while a small region was interpolated and exported as the 2D fieldmap.

\subsection{Implementation}

\begin{figure}[htbp]
  \centering
  \includegraphics[width=\linewidth]{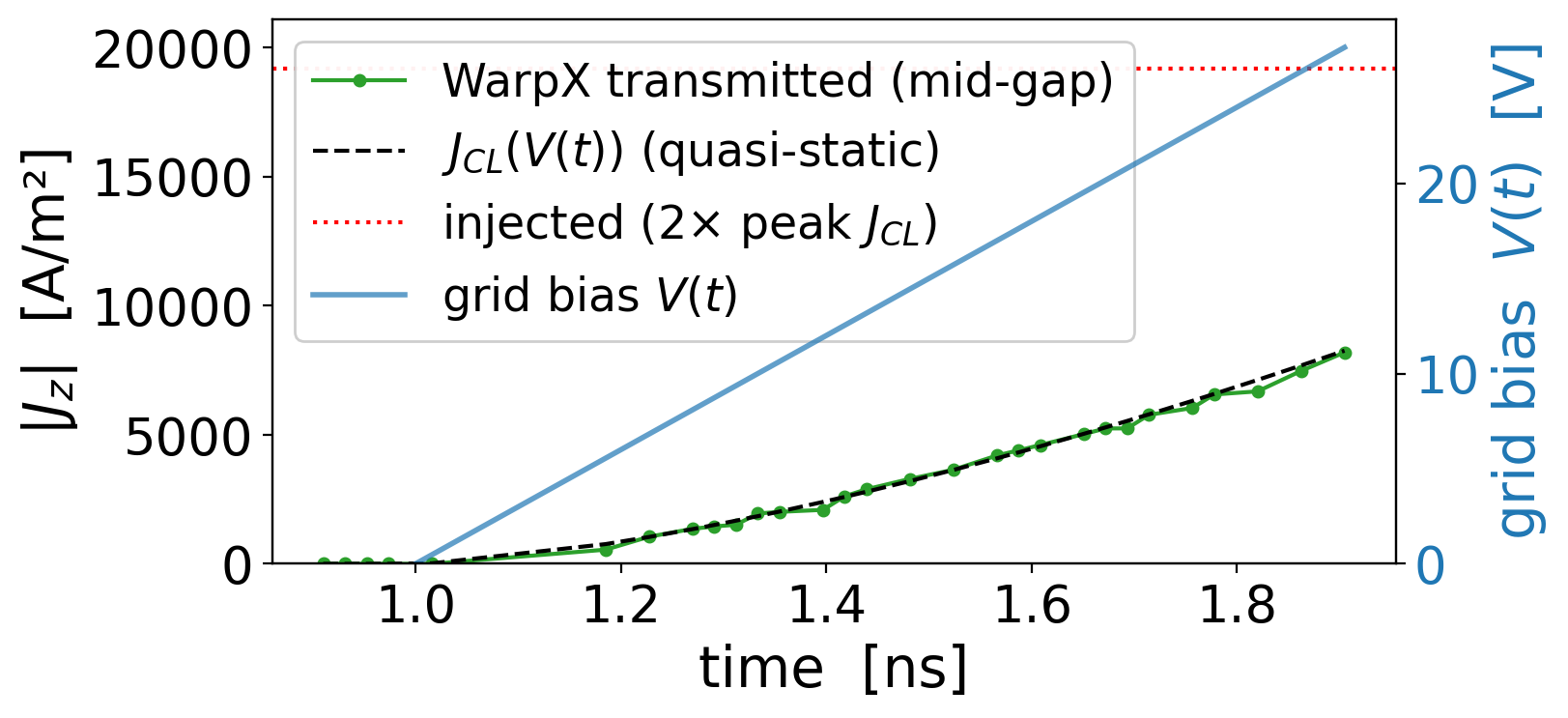}
  \caption{Over-injection of current at the cathode, reduced to the Child--Langmuir limit by space charge forces in WarpX. The analytic calculation is shown as the black dashed line, which grows as the grid voltage is increased over the 2~ns pulse.}
  \label{fig:cathode}
\end{figure}

For the emission phase, electrons are injected at twice the maximum Child--Langmuir limit. WarpX allows for a flux source, so particles are continuously emitted from the cathode at this rate. Space charge organically reduces the current density back to $J_{CL}$ to obtain the true distribution as seen in Fig.~\ref{fig:cathode}. The WarpX simulation is modeled in 2D instead of the sheet approximation.

\begin{figure}[htbp]
  \centering
  \begin{subfigure}{\linewidth}
    \centering
    \includegraphics[width=\linewidth]{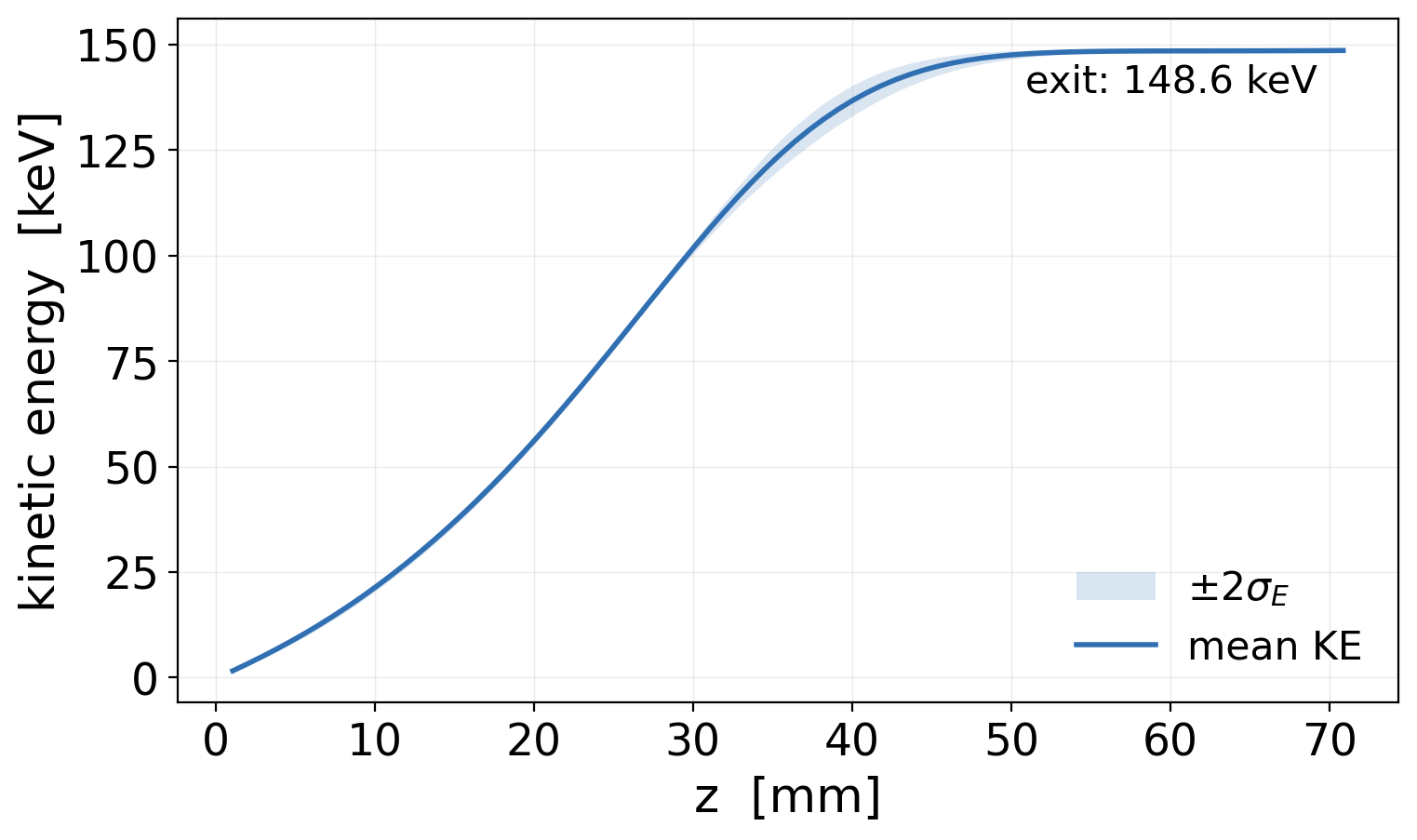}
    \caption{}
    \label{fig:gun-gain}
  \end{subfigure}
  \begin{subfigure}{\linewidth}
    \centering
    \includegraphics[width=\linewidth]{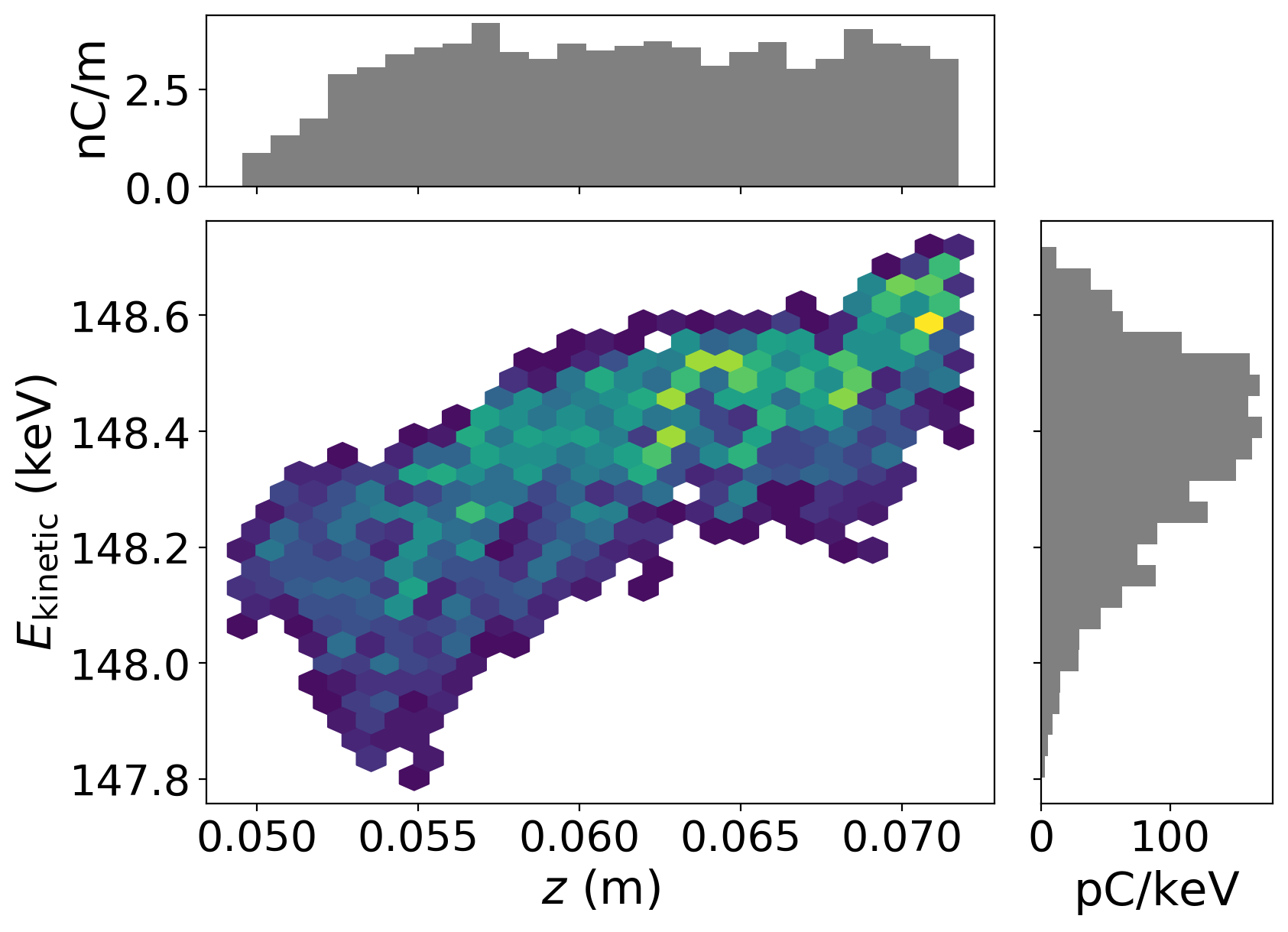}
    \caption{}
    \label{fig:gun-exit}
  \end{subfigure}
  \caption{(a) Kinetic energy gain of the electron bunch through the 150~kV CHILI electrostatic gun. (b) Longitudinal phase space of the bunch at the gun exit.}
  \label{fig:gun}
\end{figure}

The delivered flux is accelerated through the CHILI gun. As the gun field is nonuniform, most acceleration occurs in the first 50~mm, noted in Fig.~\ref{fig:gun-gain}. The longitudinal phase space of the bunch at the gun exit is shown in Fig.~\ref{fig:gun-exit}.

\section{Prebuncher}

The prebuncher consists of two rf cavities enclosed in solenoids, with the intent to compress the emitted electrons longitudinally for better capture in the downstream acceleration rf cavities. The bulk arrives at the zero crossing ($90^\circ$ off crest), while earlier electrons receive a push to slow down and later electrons are kicked forwards.

\subsection{Prebuncher Fieldmaps}
\begin{figure}[htbp]
  \centering
  \includegraphics[width=\linewidth]{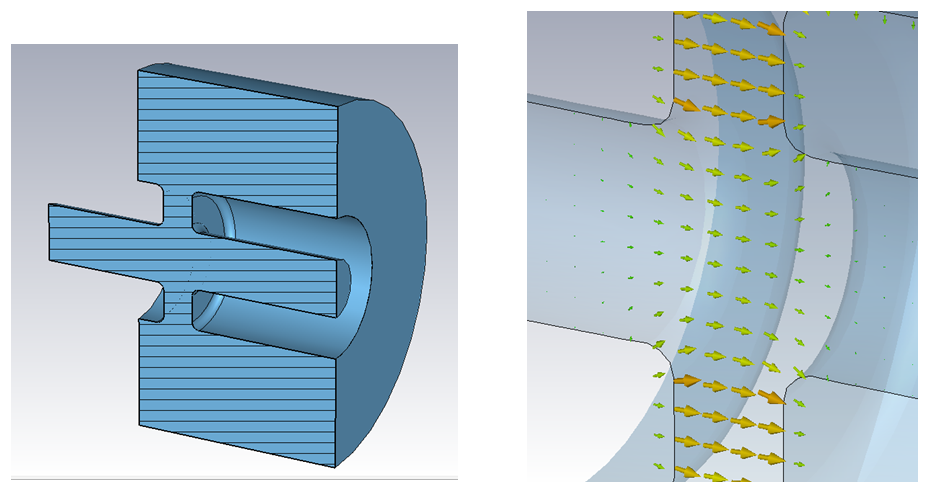}
  \caption{Microwave Studio model of the prebuncher cavity and its fundamental-mode electric field.}
  \label{fig:prebuncher-cavity}
\end{figure}

Both prebunchers share the same cavity design (though with different loaded quality factors), with the second installed in reverse. The cavities were modeled prior to this work in CST Microwave Studio with cylindrical symmetry~\cite{linacsim}. The cavity model and its fundamental-mode electric field are shown in Fig.~\ref{fig:prebuncher-cavity}, and the calculated mode frequencies are compared with measurements in Table~\ref{tab:prebuncher-modes}.

\begin{table}[htbp]
  \caption{Calculated and measured mode frequencies of the prebuncher cavity.}
  \label{tab:prebuncher-modes}
  \begin{ruledtabular}
  \begin{tabular}{ccc}
  Mode & $f_{\text{calc}}$ (MHz) & $f_{\text{meas}}$ (MHz) \\
  \hline
  1 & 215 & 214 \\
  2 & 577 & 580 \\
  3 & 614 & 643 \\
  4 & 802 & 814 \\
  5 & 834 & 843 \\
  \end{tabular}
  \end{ruledtabular}
\end{table}

The configuration file specifies the power and relative phase of each cavity, while the field amplitude is derived from physically-measured loaded quality factors: $Q = 3000$ for prebuncher~1 and $Q=4300$ for prebuncher~2~\cite{cbn942}. The prebunchers operate at 18 times the master oscillator frequency, referenced to the 1991 measurement of the CESR RF frequency (which is 42 times the master oscillator) of 499.7645~MHz~\cite{rf_numerology}.

The fieldmap is normalized to a stored energy of $U_{\text{sim}} = 1$~J, and the field amplitude is scaled by
\begin{equation}
A = \sqrt{\frac{U}{U_{\text{sim}}}}, \qquad U = \frac{Q_L P}{2\pi f},
\label{eq:prebuncher-scale}
\end{equation}
where $P$ is the dissipated power and $f$ is the operating frequency.

\subsection{Solenoids}

\begin{figure}[htbp]
  \centering
  \includegraphics[width=\linewidth]{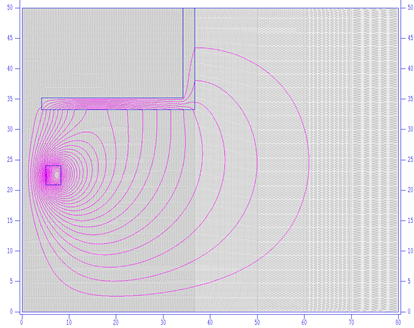}
  \caption{Simulation results for lens 0A, plotted as distance down the beam pipe (vertical) vs. radius (horizontal).}
  \label{fig:prebuncher-solenoids}
\end{figure}

The five lenses (lens 0A to 0E) and two longer solenoids (solenoid 0 and 1) were also modeled in Poisson Superfish using their schematics, including a nearby steel frame. Figure~\ref{fig:prebuncher-solenoids} shows an example simulation result for one solenoid and lens. Due to the large magnetic permeability of steel, the field lines are drawn into the frame. The sources of the field are modeled as small rectangular regions of uniform current density, with total enclosed current equal to 
\begin{equation}
  I_\text{enclosed}=\left( \#\text{ of windings} \right)\times\left( 1\text{ A} \right).
\end{equation}
Each solenoid is simulated in Superfish with 1~A of input current, and the fieldmap is scaled linearly to the operating current.

The physical injector contains cylindrically symmetric annular end plates with four equally spaced support bars. In the Superfish model, complete cylindrical symmetry is used as an approximation. This makes the model simpler and is not expected to add a significant error to the field inside of the pipe.

\subsection{Implementation}

\begin{figure}[htbp]
  \centering
  \includegraphics[width=\linewidth]{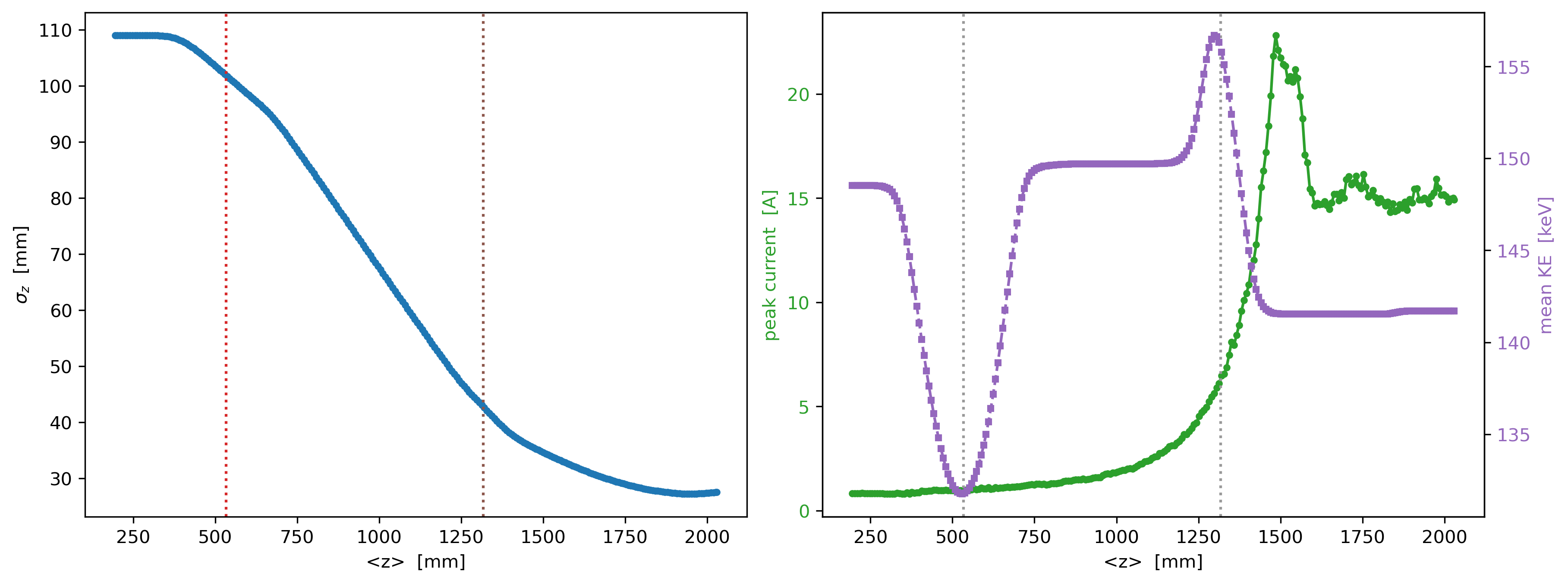}
  \caption{Left, bunch length $\sigma_z$ along the prebuncher, with the two cavity gaps marked by vertical lines. Right, peak current and mean kinetic energy.}
  \label{fig:prebuncher-line}
\end{figure}

The beam enters the prebuncher as a 1.29~nC, 2~ns pulse with an rms length of $\sigma_z = 109$~mm and mean kinetic energy of 148.5~keV. As the prebuncher cavities are set to the zero crossing, the energy change is small (142~keV at exit). Fig.~\ref{fig:prebuncher-line} shows the full beam evolution along the section, with $\sigma_z$ decreasing to 27.5~mm after compression.

\begin{figure}[htbp]
  \centering
  \includegraphics[width=\linewidth]{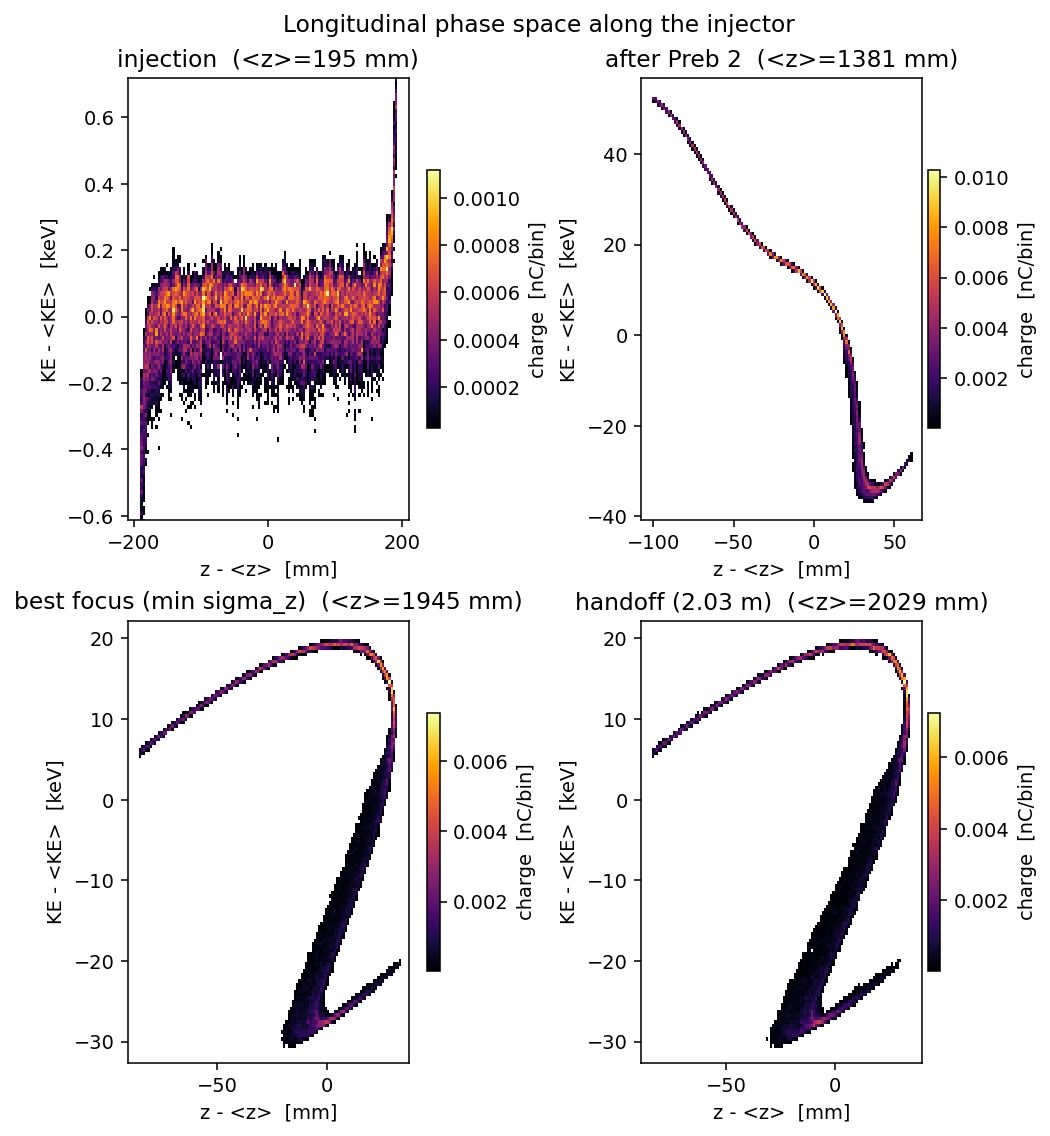}
  \caption{Longitudinal phase space at four stations along the prebuncher section: entrance, after the first cavity, after the second cavity, and at the 2.03~m handoff. The sinusoidal chirp folds the distribution.}
  \label{fig:prebuncher-phasespace}
\end{figure}

Since the 2~ns pulse is a large fraction of the 214~MHz RF period, the energy change applied by the zero crossing is not linear. As the bunch travels through the cavities, it produces the longitudinal phase space shown in Fig.~\ref{fig:prebuncher-phasespace} rather than a point.

\section{Autophasing}
\label{sec:autophasing}
For both the prebuncher and linac sections, the rf cavity configurations from the input files are specified as relative phases which are dependent on the on crest phase. This phase depends on the exact arrival time of the bunch, which can change based on the other machine settings. Therefore, during each simulation run with different parameters, the on crest phase must be rederived.

To find the crest, a sample of particles is loaded alongside the fieldmap, then the phase search is conducted using RK4 integration. This method approximately captures the energy gain without need to run the full PIC simulation. As a result, the on-crest phase can be located in a few seconds.

\begin{figure}[htbp]
  \centering
  \includegraphics[width=\linewidth]{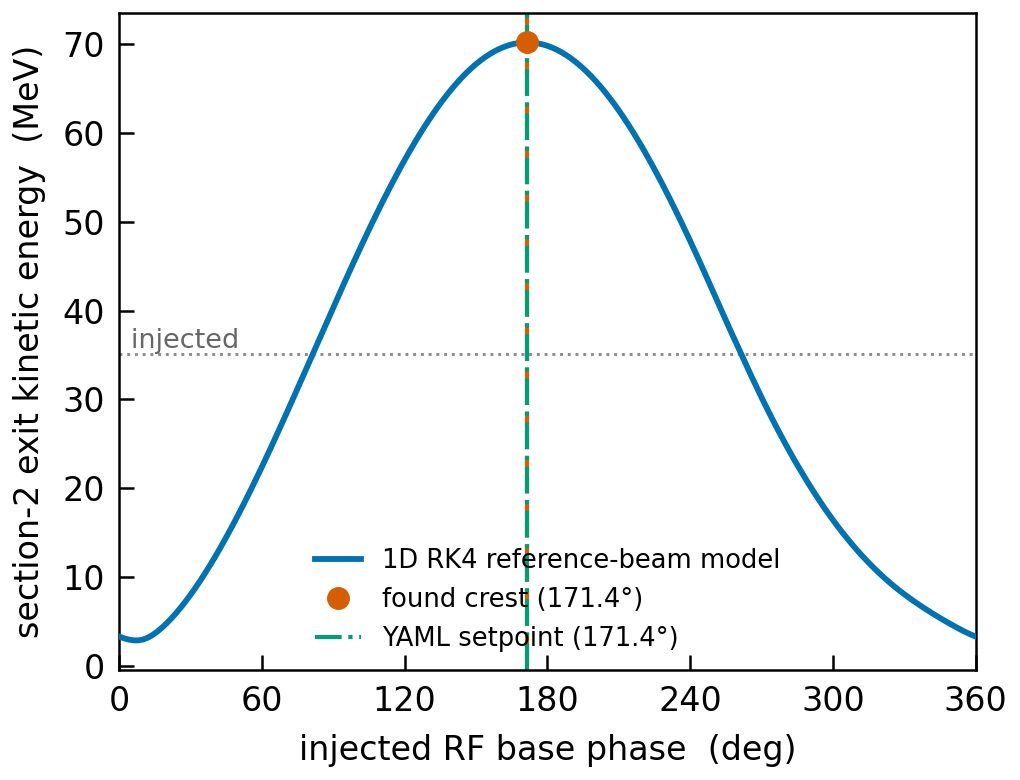}
  \caption{RK4 phase scan to find on-crest phase, which is then applied to the YAML configuration.}
  \label{fig:autophase}
\end{figure}

Coarse evaluations are first conducted in $4^\circ$ increments with 512 particles to build the scan shown in Fig.~\ref{fig:autophase}, and then a finer evaluation is conducted in a smaller range with 2048 particles and $0.1^\circ$ increments to narrow in on the on-crest phase. This is automatically saved in the configuration YAML file as \verb|CREST_PHASE_DEG|, which is then added to the offset \verb|PHASE_DEG|.

\section{Electron Linacs}
Before the positron converter, electrons are accelerated to roughly 150~MeV using four linacs. Section 1 is an exact copy of the 86-cell cavity from the two-mile SLAC accelerator. Sections 2 and 3 are 90-cell cavities from the Cambridge Electron Accelerator (CEA), and section 4 was taken from the 1966 Cornell University Linac (CU)~\cite{linacinfo}.

\subsection{Fieldmaps}
Since Section 1 is the same design from the SLAC accelerator, there is a large body of information available with specifications and simulations~\cite{twomile}.

The 3-dimensional, 86-cell cavity is too large to simulate on a typical PC with Microwave Studio, so Poisson Superfish was used instead. However, the simulation failed because Superfish is not a true eigenmode solver. Superfish internally varies the RF frequency and searches for frequencies that have a resonance. It can find normal modes, but it can easily get confused in many-cell cavity calculations with closely spaced modes. The computed field would have approximately the right appearance and symmetry, but with some obvious problems. The solution would depend strongly on the location of the field source, which should not matter.

Instead, we refer to a method found to model these cavities in the 1980s~\cite{wang1985,loew1979}. Each cell of the cavity is modeled separately, assuming perfect periodic boundary conditions. Then, the fields are stitched together, respecting the phase advance of the wave. To respect the symmetry of the cavity, 1.5 cells are modeled because the cavity is a $\frac{2\pi}{3}$ traveling wave cavity. Superfish gives a standing wave solution, which is then sampled to produce a right-going traveling wave solution. We require a constant gradient cavity to produce the boundary conditions for matching cells.

\begin{figure}[htbp]
  \centering
  \includegraphics[width=\linewidth]{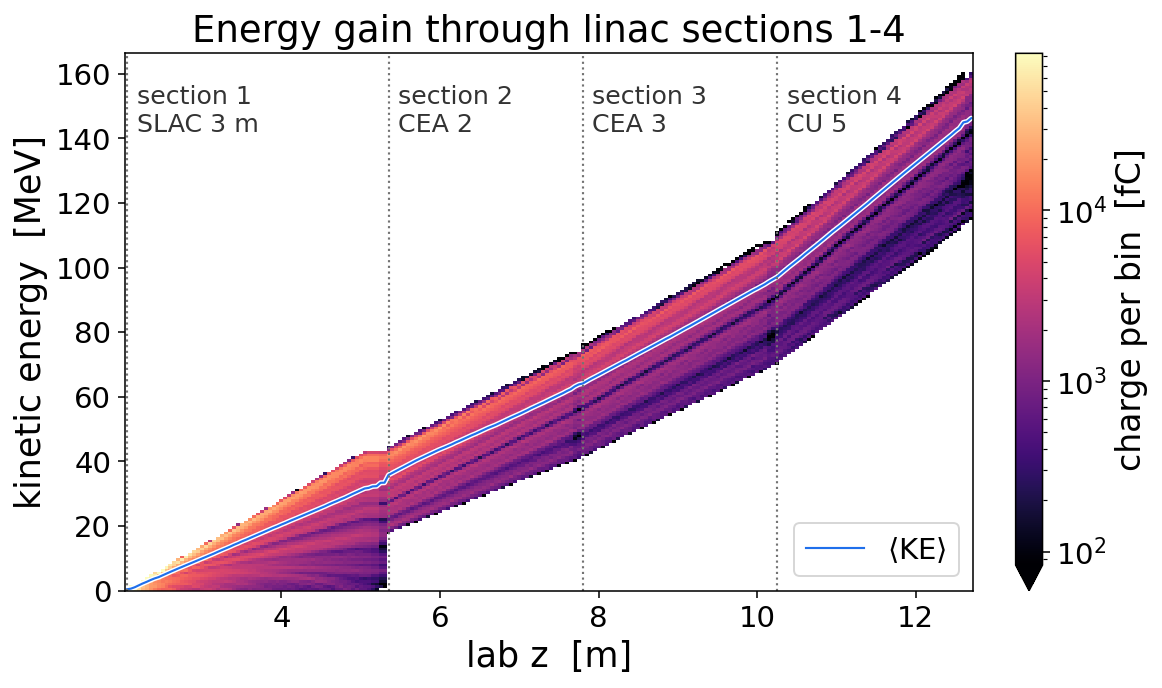}
  \caption{Kinetic energy distribution of electrons through linacs 1 to 4. Blue line represents mean energy.}
  \label{fig:electron-linac}
\end{figure}

Fieldmaps are not available for the CEA and CU linacs, however they can be inserted into the current code once produced. For now, sections 2--4 use the SLAC fieldmap which is rescaled to the appropriate gradient. This approximation must be fixed before further use. For example, the 5-meter CU linac is rescaled to a higher gradient because the SLAC linac is only 3 meters, as seen in Fig.~\ref{fig:electron-linac}.

\subsection{Implementation}

Each section is run as an independent WarpX simulation. Space charge is enabled only in section~1, where the beam is still at 150~keV; in sections 2--4 the $1/\gamma^2$ suppression at $\langle\text{KE}\rangle\ge35$~MeV makes the self-field negligible and it is disabled for computation efficiency.

The traveling wave is produced from the two Superfish fieldmaps which are loaded as separate applied fields driven at a phase offset of $\frac{\pi}{2}$. The phase is referenced to the arrival time of the bunch at the cavity entrance using the autophase method discussed in Section~\ref{sec:autophasing}.

\begin{figure}[htbp]
  \centering
  \includegraphics[width=\linewidth]{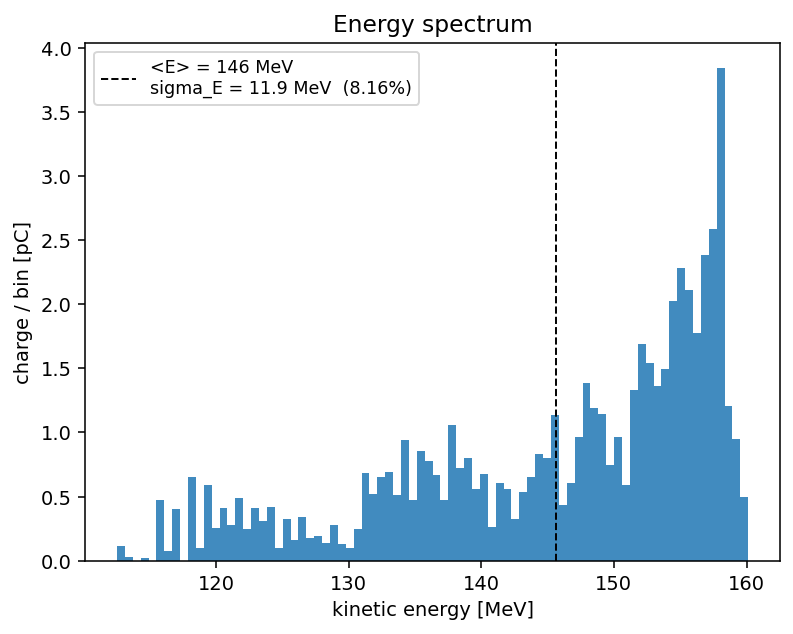}
  \caption{Output kinetic energy distribution from Section 4 of the electron linac. This is a distribution of the particles that are received by the converter.}
  \label{fig:electron-linac-out}
\end{figure}

Figure~\ref{fig:electron-linac} shows the energy spread increasing in Section 1, then remaining roughly constant through the remaining linacs. The full output distribution at the end of Section 4 is given in Fig.~\ref{fig:electron-linac-out}.

\section{Positron Converter}
The positrons are produced by a 6.35~mm tungsten target~\cite{possource}. Electrons scatter off the tungsten nuclei and radiate Bremsstrahlung photons, which pair produce and create an electromagnetic shower. The positrons from the shower are captured by a solenoid which collimates the beam.

\begin{figure}[htbp]
  \centering
  \includegraphics[width=\linewidth]{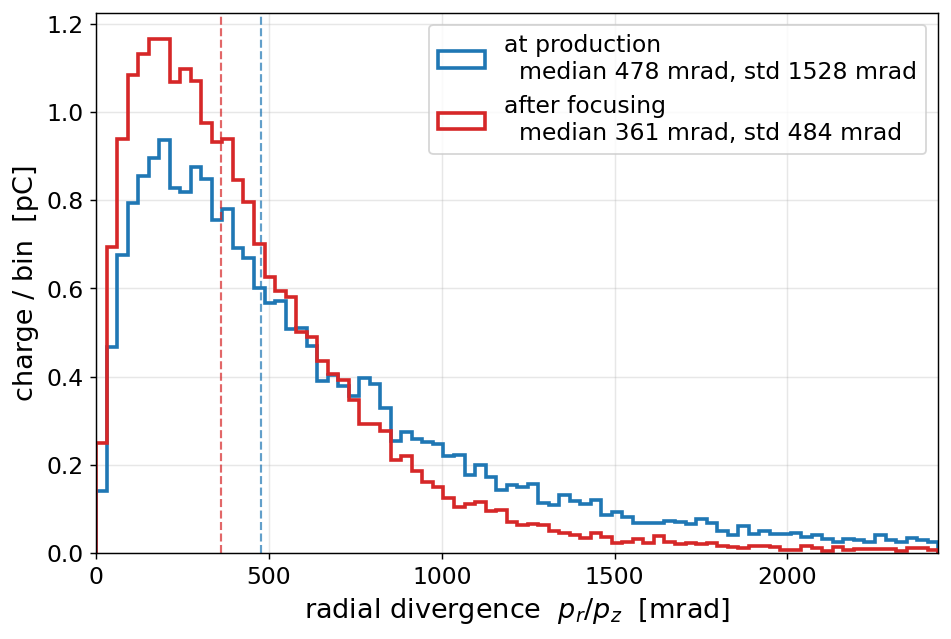}
  \caption{Positron radial divergence at production (the target back face) and after focusing
  (the section 5 entrance flange, 129~mm downstream). Divergence is invariant under a field-free
  drift, so the narrowing between the two is the capture solenoid at work.}
  \label{fig:conv-divergence}
\end{figure}

\subsection{Optics}
Figure~\ref{fig:conv-divergence} shows the initial high divergence of the post-shower positrons. These positrons are collimated using a short 0.7022~T solenoid~\cite{fromowitz}. This solenoid decreases the angular divergence from a median of 478~mrad to 361~mrad.

\subsection{Implementation}
The section 4 exit beam from WarpX is upsampled to 50,000 macroparticles, similarly to the other stages, and written as a G4beamline ASCII file with each macroparticle corresponding to an ``event.'' By tracking each event ID, output positrons are mapped back to the electron that produced them to account for charge.

The incident 59.3~pC electron beam, with an average kinetic energy of 145.7~MeV, gives a yield of 0.386 $e^+/e^-$ (as well as 1.04 $e^-/e^-$ and 11.2 $\gamma/e^-$). The 22.9~pC of positrons have a large energy spread with a mean of 17.1~MeV.

\section{Positron Linacs}
The positrons are reaccelerated by four additional cavities: two CEA and two CU linacs. These are modeled in Impact-T rather than WarpX.

As discussed previously, the post-converter positrons have a much higher energy spread and angular divergence compared to the initial electron beam, so the first two sections are enclosed in 0.243~T capture solenoids. Several quadrupole magnets also exist, primarily in sections 6--8.

\begin{figure}[htbp]
  \centering
  \includegraphics[width=\linewidth]{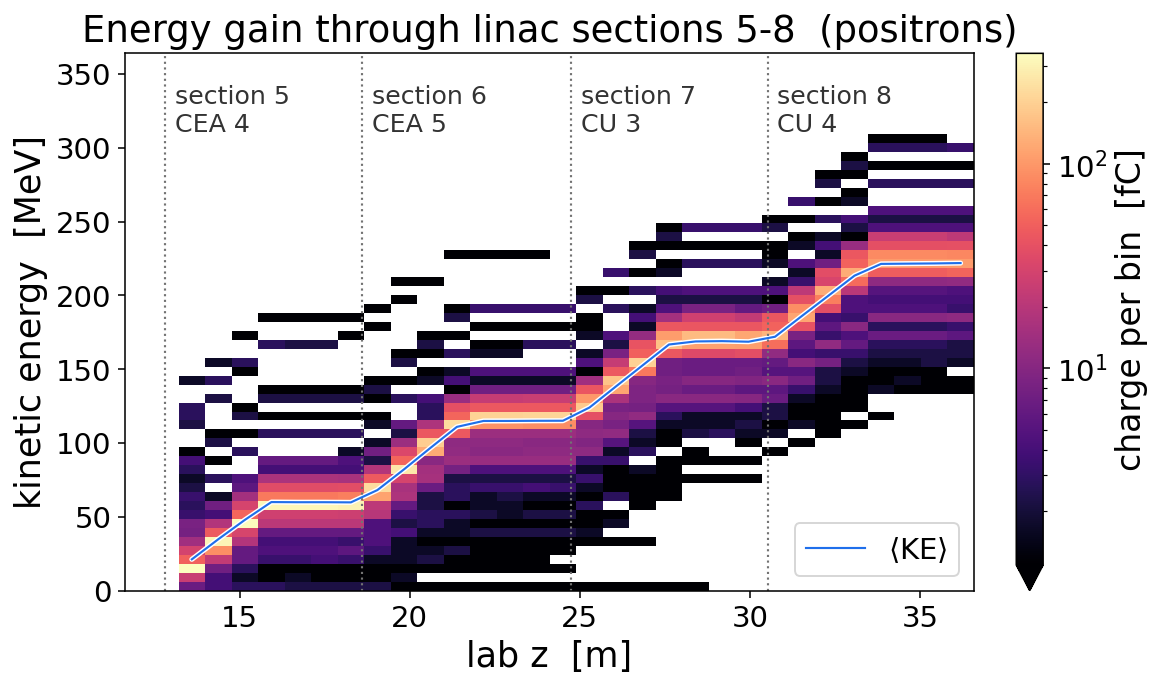}
  \caption{Kinetic energy distribution of positrons through linacs 5 to 8, with section entrances marked. Blue line represents mean energy.}
  \label{fig:positron-linac}
\end{figure}

The energy spread remains high throughout the linac, shown in Fig.~\ref{fig:positron-linac}. The core of the beam is accelerated to approximately 250~MeV.

\section{Discussion}

\begin{figure}[htbp]
  \centering
  \includegraphics[width=\linewidth]{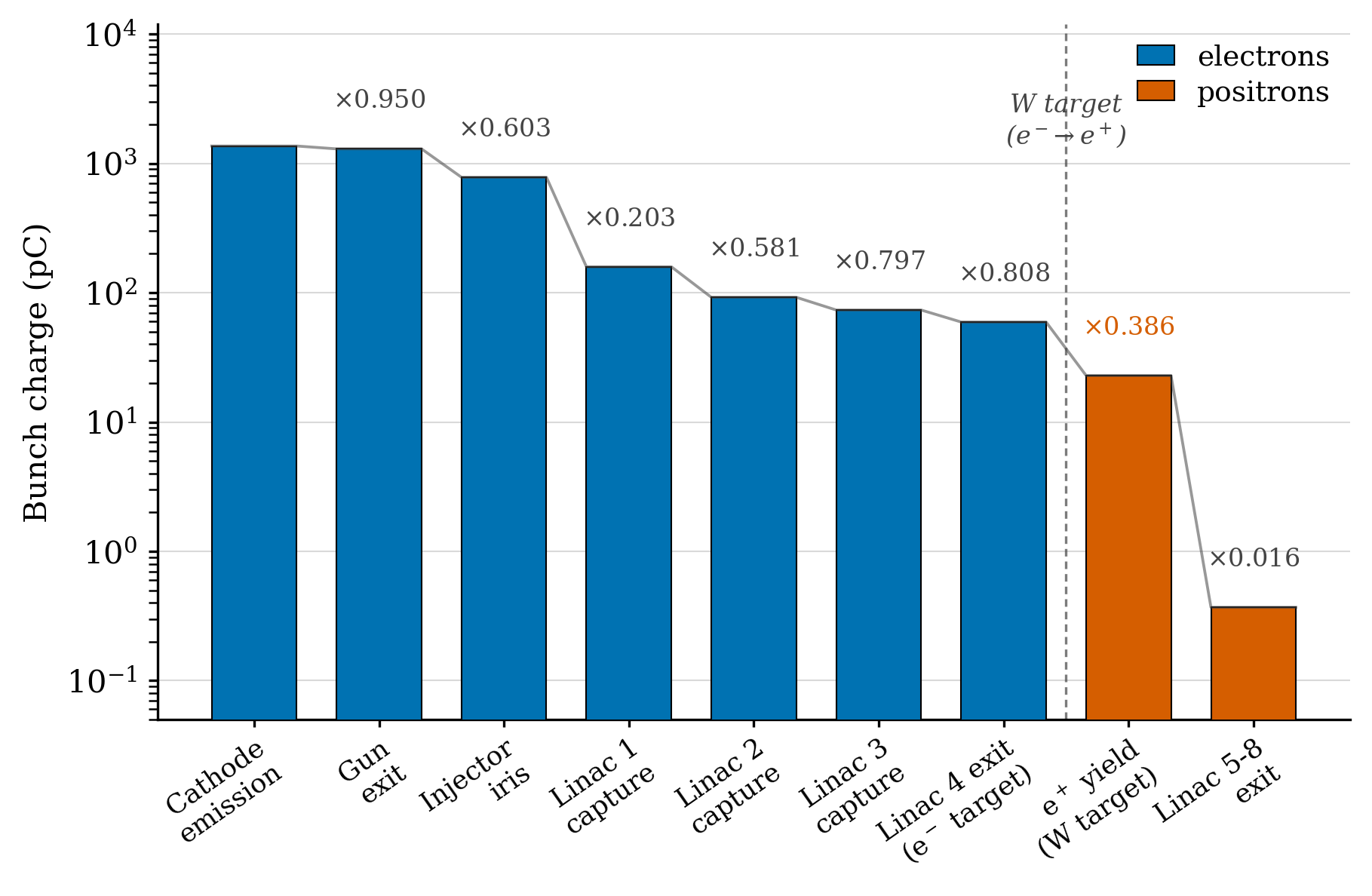}
  \caption{Amount of charge in each segment of the simulation. Only a small fraction survives to the end.}
  \label{fig:waterfall}
\end{figure}

Figure~\ref{fig:waterfall} tracks the amount of charge which survives to the end of the simulation. Of the 1.36~nC emitted at the cathode, 0.37~pC reaches the end of section 8. The largest loss comes from after the positron converter, mostly due to a) hitting the wall as a result of the large angular divergence, and b) incorrect phasing as a result of the large energy spread. Only about 1.6\% of positrons survive the linacs 5--8, in agreement with previous simulations~\cite{prevreu}.

As mentioned before, sections 2--8 reuse the SLAC fieldmap, which sets the right energy gain but are otherwise inaccurate. Future fieldmaps must be built to ensure simulation accuracy, though this will potentially require additional measurements of the existing hardware.

\subsection{Optimization}
The simulation software includes the possibility for genetic optimization using the CNSGA algorithm in Xopt~\cite{xopt}. As every stage reads settings from YAML, the optimizer can adjust the configuration without editing code. It varies 23 parameters across the chain to maximize transmitted charge while minimizing emittance, energy spread, and spot size at the exit.

Each evaluation runs in an isolated sandbox and takes roughly 40 minutes single-threaded on a consumer laptop, so evaluations can be run in parallel on single-core jobs on the CLASSE cluster. Only exploratory runs have been performed so far due to time constraints.

\subsection{Further Studies}
Previous research has looked into the potential for using rf cavities to increase the yield of linac-based slow positron sources~\cite{crisp2025}. Future work will use this software to examine the possibility of using the CESR injector for producing slow positrons, as the hardware is currently in place. However, this would require rephasing the cavities to be off-crest in order to decelerate the positrons for moderation.

\begin{acknowledgements}
This work was supported by the U.S. National Science Foundation under Award PHY-1549132, the Center for Bright Beams, as well as the NSF Research Experience for Undergraduates (REU) program.

Special thanks to previous simulations of the CESR injector linac. Prior iterations of this software were developed by Adam Bartnik, Jim Shanks, Alexandria Udenkwo, Dan Fromowitz, Kirsten Deitrick, and others.
\end{acknowledgements}

\bibliography{main}

\begin{thebibliography}{26}%
\makeatletter
\providecommand \@ifxundefined [1]{%
 \@ifx{#1\undefined}
}%
\providecommand \@ifnum [1]{%
 \ifnum #1\expandafter \@firstoftwo
 \else \expandafter \@secondoftwo
 \fi
}%
\providecommand \@ifx [1]{%
 \ifx #1\expandafter \@firstoftwo
 \else \expandafter \@secondoftwo
 \fi
}%
\providecommand \natexlab [1]{#1}%
\providecommand \enquote  [1]{``#1''}%
\providecommand \bibnamefont  [1]{#1}%
\providecommand \bibfnamefont [1]{#1}%
\providecommand \citenamefont [1]{#1}%
\providecommand \href@noop [0]{\@secondoftwo}%
\providecommand \href [0]{\begingroup \@sanitize@url \@href}%
\providecommand \@href[1]{\@@startlink{#1}\@@href}%
\providecommand \@@href[1]{\endgroup#1\@@endlink}%
\providecommand \@sanitize@url [0]{\catcode `\\12\catcode `\$12\catcode `\&12\catcode `\#12\catcode `\^12\catcode `\_12\catcode `\%12\relax}%
\providecommand \@@startlink[1]{}%
\providecommand \@@endlink[0]{}%
\providecommand \url  [0]{\begingroup\@sanitize@url \@url }%
\providecommand \@url [1]{\endgroup\@href {#1}{\urlprefix }}%
\providecommand \urlprefix  [0]{URL }%
\providecommand \Eprint [0]{\href }%
\providecommand \doibase [0]{https://doi.org/}%
\providecommand \selectlanguage [0]{\@gobble}%
\providecommand \bibinfo  [0]{\@secondoftwo}%
\providecommand \bibfield  [0]{\@secondoftwo}%
\providecommand \translation [1]{[#1]}%
\providecommand \BibitemOpen [0]{}%
\providecommand \bibitemStop [0]{}%
\providecommand \bibitemNoStop [0]{.\EOS\space}%
\providecommand \EOS [0]{\spacefactor3000\relax}%
\providecommand \BibitemShut  [1]{\csname bibitem#1\endcsname}%
\let\auto@bib@innerbib\@empty
\bibitem [{\citenamefont {Shanks}\ \emph {et~al.}(2019)\citenamefont {Shanks}, \citenamefont {Barley}, \citenamefont {Barrett}, \citenamefont {Billing}, \citenamefont {Codner}, \citenamefont {Li}, \citenamefont {Liu}, \citenamefont {Lyndaker}, \citenamefont {Rice}, \citenamefont {Rider}, \citenamefont {Rubin}, \citenamefont {Temnykh},\ and\ \citenamefont {Wang}}]{chessu}%
  \BibitemOpen
  \bibfield  {author} {\bibinfo {author} {\bibfnamefont {J.}~\bibnamefont {Shanks}}, \bibinfo {author} {\bibfnamefont {J.}~\bibnamefont {Barley}}, \bibinfo {author} {\bibfnamefont {S.}~\bibnamefont {Barrett}}, \bibinfo {author} {\bibfnamefont {M.}~\bibnamefont {Billing}}, \bibinfo {author} {\bibfnamefont {G.}~\bibnamefont {Codner}}, \bibinfo {author} {\bibfnamefont {Y.}~\bibnamefont {Li}}, \bibinfo {author} {\bibfnamefont {X.}~\bibnamefont {Liu}}, \bibinfo {author} {\bibfnamefont {A.}~\bibnamefont {Lyndaker}}, \bibinfo {author} {\bibfnamefont {D.}~\bibnamefont {Rice}}, \bibinfo {author} {\bibfnamefont {N.}~\bibnamefont {Rider}}, \bibinfo {author} {\bibfnamefont {D.~L.}\ \bibnamefont {Rubin}}, \bibinfo {author} {\bibfnamefont {A.}~\bibnamefont {Temnykh}},\ and\ \bibinfo {author} {\bibfnamefont {S.~T.}\ \bibnamefont {Wang}},\ }\bibfield  {title} {\bibinfo {title} {Accelerator design for the {Cornell} high energy synchrotron source upgrade},\ }\href {https://doi.org/10.1103/PhysRevAccelBeams.22.021602} {\bibfield  {journal} {\bibinfo  {journal} {Phys. Rev. Accel. Beams}\ }\textbf {\bibinfo {volume} {22}},\ \bibinfo {pages} {021602} (\bibinfo {year} {2019})}\BibitemShut {NoStop}%
\bibitem [{\citenamefont {Bartnik}(2013)}]{linacsim}%
  \BibitemOpen
  \bibfield  {author} {\bibinfo {author} {\bibfnamefont {A.~C.}\ \bibnamefont {Bartnik}},\ }\href {https://cesrwww.lepp.cornell.edu/wiki/CESR/LinacSimDetails} {\bibinfo {title} {{LinacSim} details}},\ \bibinfo {howpublished} {{CESR} Wiki, Cornell Laboratory for Accelerator-based Sciences and Education ({CLASSE})} (\bibinfo {year} {2013})\BibitemShut {NoStop}%
\bibitem [{\citenamefont {van~der Geer}\ and\ \citenamefont {de~Loos}(1996)}]{gpt}%
  \BibitemOpen
  \bibfield  {author} {\bibinfo {author} {\bibfnamefont {S.~B.}\ \bibnamefont {van~der Geer}}\ and\ \bibinfo {author} {\bibfnamefont {M.~J.}\ \bibnamefont {de~Loos}},\ }\bibfield  {title} {\bibinfo {title} {General particle tracer: A new 3d code for accelerator and beamline design},\ }in\ \href@noop {} {\emph {\bibinfo {booktitle} {Proceedings of the 5th European Particle Accelerator Conference (EPAC'96)}}}\ (\bibinfo {address} {Sitges, Spain},\ \bibinfo {year} {1996})\ pp.\ \bibinfo {pages} {1241--1243}\BibitemShut {NoStop}%
\bibitem [{\citenamefont {Sagan}(2006)}]{bmad}%
  \BibitemOpen
  \bibfield  {author} {\bibinfo {author} {\bibfnamefont {D.}~\bibnamefont {Sagan}},\ }\bibfield  {title} {\bibinfo {title} {Bmad: A relativistic charged particle simulation library},\ }\href {https://doi.org/10.1016/j.nima.2005.11.001} {\bibfield  {journal} {\bibinfo  {journal} {Nuclear Instruments and Methods in Physics Research Section A}\ }\textbf {\bibinfo {volume} {558}},\ \bibinfo {pages} {356} (\bibinfo {year} {2006})}\BibitemShut {NoStop}%
\bibitem [{\citenamefont {Vay}\ \emph {et~al.}(2018)\citenamefont {Vay}, \citenamefont {Grote}, \citenamefont {Lehe}, \citenamefont {Chang} \emph {et~al.}}]{warpx}%
  \BibitemOpen
  \bibfield  {author} {\bibinfo {author} {\bibfnamefont {J.-L.}\ \bibnamefont {Vay}}, \bibinfo {author} {\bibfnamefont {D.~P.}\ \bibnamefont {Grote}}, \bibinfo {author} {\bibfnamefont {R.}~\bibnamefont {Lehe}}, \bibinfo {author} {\bibfnamefont {H.-S.}\ \bibnamefont {Chang}}, \emph {et~al.},\ }\bibfield  {title} {\bibinfo {title} {Warpx: A new exascale computing platform for beam--plasma simulations},\ }\href {https://doi.org/10.1016/j.nima.2018.01.035} {\bibfield  {journal} {\bibinfo  {journal} {Nuclear Instruments and Methods in Physics Research Section A}\ }\textbf {\bibinfo {volume} {909}},\ \bibinfo {pages} {476} (\bibinfo {year} {2018})}\BibitemShut {NoStop}%
\bibitem [{\citenamefont {Agostinelli}\ \emph {et~al.}(2003)\citenamefont {Agostinelli} \emph {et~al.}}]{geant4}%
  \BibitemOpen
  \bibfield  {author} {\bibinfo {author} {\bibfnamefont {S.}~\bibnamefont {Agostinelli}} \emph {et~al.},\ }\bibfield  {title} {\bibinfo {title} {Geant4---a simulation toolkit},\ }\href {https://doi.org/10.1016/S0168-9002(03)01368-8} {\bibfield  {journal} {\bibinfo  {journal} {Nuclear Instruments and Methods in Physics Research Section A}\ }\textbf {\bibinfo {volume} {506}},\ \bibinfo {pages} {250} (\bibinfo {year} {2003})}\BibitemShut {NoStop}%
\bibitem [{\citenamefont {Qiang}\ \emph {et~al.}(2000)\citenamefont {Qiang}, \citenamefont {Ryne}, \citenamefont {Habib},\ and\ \citenamefont {Decyk}}]{impactt}%
  \BibitemOpen
  \bibfield  {author} {\bibinfo {author} {\bibfnamefont {J.}~\bibnamefont {Qiang}}, \bibinfo {author} {\bibfnamefont {R.~D.}\ \bibnamefont {Ryne}}, \bibinfo {author} {\bibfnamefont {S.}~\bibnamefont {Habib}},\ and\ \bibinfo {author} {\bibfnamefont {V.}~\bibnamefont {Decyk}},\ }\bibfield  {title} {\bibinfo {title} {An object-oriented parallel particle-in-cell code for beam dynamics simulation in linear accelerators},\ }\href {https://doi.org/10.1006/jcph.2000.6570} {\bibfield  {journal} {\bibinfo  {journal} {Journal of Computational Physics}\ }\textbf {\bibinfo {volume} {163}},\ \bibinfo {pages} {434} (\bibinfo {year} {2000})}\BibitemShut {NoStop}%
\bibitem [{\citenamefont {Mayes}\ \emph {et~al.}(2024)\citenamefont {Mayes} \emph {et~al.}}]{xopt}%
  \BibitemOpen
  \bibfield  {author} {\bibinfo {author} {\bibfnamefont {C.~E.}\ \bibnamefont {Mayes}} \emph {et~al.},\ }\bibfield  {title} {\bibinfo {title} {Xopt: A generator for autonomous experimentation and black-box optimization},\ }\href {https://doi.org/10.21105/joss.05993} {\bibfield  {journal} {\bibinfo  {journal} {Journal of Open Source Software}\ }\textbf {\bibinfo {volume} {9}},\ \bibinfo {pages} {5993} (\bibinfo {year} {2024})}\BibitemShut {NoStop}%
\bibitem [{\citenamefont {Roberts}(2009)}]{g4beamline}%
  \BibitemOpen
  \bibfield  {author} {\bibinfo {author} {\bibfnamefont {T.}~\bibnamefont {Roberts}},\ }\bibfield  {title} {\bibinfo {title} {{G4beamline} particle tracking in matter-dominated beam lines},\ }\href {https://doi.org/10.1063/1.3120075} {\bibfield  {journal} {\bibinfo  {journal} {AIP Conference Proceedings}\ }\textbf {\bibinfo {volume} {1099}},\ \bibinfo {pages} {440} (\bibinfo {year} {2009})}\BibitemShut {NoStop}%
\bibitem [{\citenamefont {Billen}\ and\ \citenamefont {Young}(2004)}]{superfish}%
  \BibitemOpen
  \bibfield  {author} {\bibinfo {author} {\bibfnamefont {J.~H.}\ \bibnamefont {Billen}}\ and\ \bibinfo {author} {\bibfnamefont {L.~M.}\ \bibnamefont {Young}},\ }\href@noop {} {\bibinfo {title} {{POISSON SUPERFISH}}} (\bibinfo {year} {2004})\BibitemShut {NoStop}%
\bibitem [{\citenamefont {Fromowitz}(2000)}]{fromowitz}%
  \BibitemOpen
  \bibfield  {author} {\bibinfo {author} {\bibfnamefont {D.~B.}\ \bibnamefont {Fromowitz}},\ }\emph {\bibinfo {title} {Increasing the Positron Capture Efficiency of the {CESR} Linac Injector}},\ \href@noop {} {Ph.D. thesis},\ \bibinfo  {school} {Cornell University}, \bibinfo {address} {Ithaca, New York} (\bibinfo {year} {2000})\BibitemShut {NoStop}%
\bibitem [{\citenamefont {Huebl}\ \emph {et~al.}(2024)\citenamefont {Huebl}, \citenamefont {Lehe}, \citenamefont {Pausch} \emph {et~al.}}]{openpmd}%
  \BibitemOpen
  \bibfield  {author} {\bibinfo {author} {\bibfnamefont {A.}~\bibnamefont {Huebl}}, \bibinfo {author} {\bibfnamefont {R.}~\bibnamefont {Lehe}}, \bibinfo {author} {\bibfnamefont {R.}~\bibnamefont {Pausch}}, \emph {et~al.},\ }\bibfield  {title} {\bibinfo {title} {openpmd: A meta data standard for particle and mesh-based data},\ }\href {https://doi.org/10.1016/j.cpc.2023.108908} {\bibfield  {journal} {\bibinfo  {journal} {Computer Physics Communications}\ }\textbf {\bibinfo {volume} {294}},\ \bibinfo {pages} {108908} (\bibinfo {year} {2024})}\BibitemShut {NoStop}%
\bibitem [{\citenamefont {Blum}\ \emph {et~al.}(1983)\citenamefont {Blum}, \citenamefont {Billing}, \citenamefont {Dunnam}, \citenamefont {Littauer}, \citenamefont {McDaniel}, \citenamefont {Rice}, \citenamefont {Sakazaki},\ and\ \citenamefont {Siemann}}]{chili}%
  \BibitemOpen
  \bibfield  {author} {\bibinfo {author} {\bibfnamefont {E.~B.}\ \bibnamefont {Blum}}, \bibinfo {author} {\bibfnamefont {M.~G.}\ \bibnamefont {Billing}}, \bibinfo {author} {\bibfnamefont {C.~R.}\ \bibnamefont {Dunnam}}, \bibinfo {author} {\bibfnamefont {R.~M.}\ \bibnamefont {Littauer}}, \bibinfo {author} {\bibfnamefont {B.~D.}\ \bibnamefont {McDaniel}}, \bibinfo {author} {\bibfnamefont {D.~H.}\ \bibnamefont {Rice}}, \bibinfo {author} {\bibfnamefont {L.~E.}\ \bibnamefont {Sakazaki}},\ and\ \bibinfo {author} {\bibfnamefont {R.~H.}\ \bibnamefont {Siemann}},\ }\bibfield  {title} {\bibinfo {title} {Performance of the {Cornell} high intensity linac injector},\ }in\ \href@noop {} {\emph {\bibinfo {booktitle} {Proceedings of the 12th International Conference on High-Energy Accelerators}}}\ (\bibinfo {address} {Batavia, Illinois},\ \bibinfo {year} {1983})\ pp.\ \bibinfo {pages} {262--264}\BibitemShut {NoStop}%
\bibitem [{\citenamefont {Billing}(2000)}]{billing2000}%
  \BibitemOpen
  \bibfield  {author} {\bibinfo {author} {\bibfnamefont {M.}~\bibnamefont {Billing}},\ }\href@noop {} {\bibinfo {title} {Linac injector}},\ \bibinfo {howpublished} {Internal note, Laboratory of Nuclear Studies, Cornell University} (\bibinfo {year} {2000}),\ \bibinfo {note} {dated 6 November 2000}\BibitemShut {NoStop}%
\bibitem [{\citenamefont {Child}(1911)}]{child1911}%
  \BibitemOpen
  \bibfield  {author} {\bibinfo {author} {\bibfnamefont {C.~D.}\ \bibnamefont {Child}},\ }\bibfield  {title} {\bibinfo {title} {Discharge from hot {CaO}},\ }\href {https://doi.org/10.1103/PhysRevSeriesI.32.492} {\bibfield  {journal} {\bibinfo  {journal} {Phys. Rev. (Series I)}\ }\textbf {\bibinfo {volume} {32}},\ \bibinfo {pages} {492} (\bibinfo {year} {1911})}\BibitemShut {NoStop}%
\bibitem [{\citenamefont {Langmuir}(1913)}]{langmuir1913}%
  \BibitemOpen
  \bibfield  {author} {\bibinfo {author} {\bibfnamefont {I.}~\bibnamefont {Langmuir}},\ }\bibfield  {title} {\bibinfo {title} {The effect of space charge and residual gases on thermionic currents in high vacuum},\ }\href {https://doi.org/10.1103/PhysRev.2.450} {\bibfield  {journal} {\bibinfo  {journal} {Phys. Rev.}\ }\textbf {\bibinfo {volume} {2}},\ \bibinfo {pages} {450} (\bibinfo {year} {1913})}\BibitemShut {NoStop}%
\bibitem [{\citenamefont {{Cornell Laboratory for Elementary-Particle Physics}}(2009)}]{chiligun_dwg}%
  \BibitemOpen
  \bibfield  {author} {\bibinfo {author} {\bibnamefont {{Cornell Laboratory for Elementary-Particle Physics}}},\ }\href@noop {} {\bibinfo {title} {{CHILI} gun {MkII} assembly}},\ \bibinfo {howpublished} {Engineering drawing 6051-022 rev.~B, sheet 1, Floyd R. Newman Laboratory, Cornell University} (\bibinfo {year} {2009})\BibitemShut {NoStop}%
\bibitem [{\citenamefont {Kazacha}(1994)}]{cbn942}%
  \BibitemOpen
  \bibfield  {author} {\bibinfo {author} {\bibfnamefont {V.}~\bibnamefont {Kazacha}},\ }\href@noop {} {\bibinfo {title} {Calibration of the prebuncher cavity accelerating voltages}},\ \bibinfo {howpublished} {{CBN} 94-2, Laboratory of Nuclear Studies, Cornell University} (\bibinfo {year} {1994})\BibitemShut {NoStop}%
\bibitem [{\citenamefont {Giannella}\ and\ \citenamefont {Barley}(1992)}]{rf_numerology}%
  \BibitemOpen
  \bibfield  {author} {\bibinfo {author} {\bibfnamefont {M.}~\bibnamefont {Giannella}}\ and\ \bibinfo {author} {\bibfnamefont {J.}~\bibnamefont {Barley}},\ }\href {https://cesrwww.classe.cornell.edu/documents/control/hardware/fast_timing/rf_numerology.html} {\bibinfo {title} {{RF} numerology}},\ \bibinfo {howpublished} {{CESR} operations document, Cornell Laboratory for Accelerator-based Sciences and Education ({CLASSE})} (\bibinfo {year} {1992})\BibitemShut {NoStop}%
\bibitem [{\citenamefont {Codner}(2022)}]{linacinfo}%
  \BibitemOpen
  \bibfield  {author} {\bibinfo {author} {\bibfnamefont {G.}~\bibnamefont {Codner}},\ }\href {https://cesrwww.lepp.cornell.edu/wiki/CESR/LinacHistory} {\bibinfo {title} {{CESR Linac} current configuration and history}},\ \bibinfo {howpublished} {{CESR} Wiki, Cornell Laboratory for Accelerator-based Sciences and Education ({CLASSE})} (\bibinfo {year} {2022})\BibitemShut {NoStop}%
\bibitem [{\citenamefont {Neal}\ \emph {et~al.}(1968)\citenamefont {Neal}, \citenamefont {Dupen}, \citenamefont {Hogg},\ and\ \citenamefont {Loew}}]{twomile}%
  \BibitemOpen
  \bibfield  {author} {\bibinfo {author} {\bibfnamefont {R.~B.}\ \bibnamefont {Neal}}, \bibinfo {author} {\bibfnamefont {D.~W.}\ \bibnamefont {Dupen}}, \bibinfo {author} {\bibfnamefont {H.~A.}\ \bibnamefont {Hogg}},\ and\ \bibinfo {author} {\bibfnamefont {G.~A.}\ \bibnamefont {Loew}},\ }\href@noop {} {\emph {\bibinfo {title} {The {Stanford} Two-Mile Accelerator}}},\ edited by\ \bibinfo {editor} {\bibfnamefont {R.~B.}\ \bibnamefont {Neal}}\ (\bibinfo  {publisher} {W. A. Benjamin, Inc.},\ \bibinfo {address} {New York},\ \bibinfo {year} {1968})\BibitemShut {NoStop}%
\bibitem [{\citenamefont {Wang}\ and\ \citenamefont {Loew}(1985)}]{wang1985}%
  \BibitemOpen
  \bibfield  {author} {\bibinfo {author} {\bibfnamefont {J.~W.}\ \bibnamefont {Wang}}\ and\ \bibinfo {author} {\bibfnamefont {G.~A.}\ \bibnamefont {Loew}},\ }\href@noop {} {\emph {\bibinfo {title} {Measurements of Ultimate Accelerating Gradients in the {SLAC} Disk-Loaded Structure (Part~I)}}},\ \bibinfo {type} {Tech. Rep.}\ \bibinfo {number} {{SLAC/AP-26}}\ (\bibinfo  {institution} {Stanford Linear Accelerator Center, Stanford University},\ \bibinfo {year} {1985})\BibitemShut {NoStop}%
\bibitem [{\citenamefont {Loew}\ \emph {et~al.}(1979)\citenamefont {Loew}, \citenamefont {Miller}, \citenamefont {Early},\ and\ \citenamefont {Bane}}]{loew1979}%
  \BibitemOpen
  \bibfield  {author} {\bibinfo {author} {\bibfnamefont {G.~A.}\ \bibnamefont {Loew}}, \bibinfo {author} {\bibfnamefont {R.~H.}\ \bibnamefont {Miller}}, \bibinfo {author} {\bibfnamefont {R.~A.}\ \bibnamefont {Early}},\ and\ \bibinfo {author} {\bibfnamefont {K.~L.}\ \bibnamefont {Bane}},\ }\bibfield  {title} {\bibinfo {title} {Computer calculations of traveling-wave periodic structure properties},\ }in\ \href@noop {} {\emph {\bibinfo {booktitle} {Proceedings of the 1979 Particle Accelerator Conference}}}\ (\bibinfo {address} {San Francisco, California},\ \bibinfo {year} {1979})\BibitemShut {NoStop}%
\bibitem [{\citenamefont {Mikhailichenko}(2015)}]{possource}%
  \BibitemOpen
  \bibfield  {author} {\bibinfo {author} {\bibfnamefont {A.}~\bibnamefont {Mikhailichenko}},\ }\href {https://arxiv.org/abs/1505.01406} {\bibinfo {title} {{CESR}'s positron source}},\ \bibinfo {howpublished} {Presented at the {SLAC} Linear Collider Workshop ({LC02}), Stanford, California, 5 February 2002} (\bibinfo {year} {2015}),\ \bibinfo {note} {{arXiv:1505.01406}},\ \Eprint {https://arxiv.org/abs/1505.01406} {arXiv:1505.01406 [physics.acc-ph]} \BibitemShut {NoStop}%
\bibitem [{\citenamefont {Udenkwo}(2019)}]{prevreu}%
  \BibitemOpen
  \bibfield  {author} {\bibinfo {author} {\bibfnamefont {A.}~\bibnamefont {Udenkwo}},\ }\href {https://cornell.app.box.com/s/jyldcbuo0ukbyfde0kpv55w5qhmp1366} {\bibinfo {title} {Modeling the cornell electron storage ring's linear accelerator}},\ \bibinfo {howpublished} {CLASSE REU Report} (\bibinfo {year} {2019})\BibitemShut {NoStop}%
\bibitem [{\citenamefont {Crisp}\ \emph {et~al.}(2025)\citenamefont {Crisp}, \citenamefont {Goldman}, \citenamefont {Ismail},\ and\ \citenamefont {Gessner}}]{crisp2025}%
  \BibitemOpen
  \bibfield  {author} {\bibinfo {author} {\bibfnamefont {S.}~\bibnamefont {Crisp}}, \bibinfo {author} {\bibfnamefont {R.}~\bibnamefont {Goldman}}, \bibinfo {author} {\bibfnamefont {A.}~\bibnamefont {Ismail}},\ and\ \bibinfo {author} {\bibfnamefont {S.}~\bibnamefont {Gessner}},\ }\bibfield  {title} {\bibinfo {title} {Start-to-end simulations of a compact, linac-based positron source},\ }in\ \href {https://doi.org/10.18429/JACoW-NAPAC2025-WEP085} {\emph {\bibinfo {booktitle} {Proc. 6th North American Particle Accelerator Conference ({NAPAC2025})}}}\ (\bibinfo {address} {Sacramento, California},\ \bibinfo {year} {2025})\ pp.\ \bibinfo {pages} {862--865}\BibitemShut {NoStop}%
\end{thebibliography}%

\end{document}